\documentstyle[12pt]{article}
\begin{document}

\begin{center}
{\bf A Precise Measurement of the Weak Mixing Angle\\
in Neutrino-Nucleon Scattering}
\end{center}

\vskip .2in

\parskip 0.12in
\centerline{
C.G. Arroyo, B.J. King, K.T. Bachmann,
\footnote{Present address: National Center for Atmospheric
Research,
Boulder, CO 80307}
A.O. Bazarko, T. Bolton, C. Foudas,
\footnote{Present address: University of Wisconsin, Madison, WI
53706.}
}
\centerline{
 W.C. Lefmann, W.C. Leung,
S.R. Mishra,\footnote{Present address: Harvard University,
Cambridge, MA 02138.}
E. Oltman, \footnote{Present address: Lawrence Berkeley
Laboratory, Berkeley, CA 94720.}
P.Z. Quintas,\footnote{Present address: Fermilab, Batavia, IL
60510.}
}
\centerline{ S.A. Rabinowitz,
F.J. Sciulli, W.G. Seligman, M.H. Shaevitz
}
\centerline{\bf Columbia University, New York, NY 10027}

\centerline{
F.S. Merritt, M.J. Oreglia,
B.A. Schumm,$^5$
}
\centerline{\bf University of Chicago, Chicago, IL 60637}

\centerline{ R.H. Bernstein, F. Borcherding, H.E. Fisk, M.J. Lamm,}
\centerline{
W. Marsh, K.W.B. Merritt, H.M. Schellman,
\footnote{Present address: Northwestern University, Evanston, IL
60208.}
D.D. Yovanovitch
}
\centerline{\bf Fermilab, Batavia, IL 60510}

\centerline{
A. Bodek, H.S. Budd,
P. de Barbaro, W.K. Sakumoto
}
\centerline{\bf University of Rochester, Rochester, NY 14627}

\centerline{
T. Kinnel, P.H. Sandler, \footnote{Present address: Lawrence
Livermore National Laboratory, Livermore, CA  94550.}
W.H. Smith
}
\centerline{\bf University of Wisconsin, Madison, WI 53706.}

\newpage

{\hspace*{0.3cm}
        We report a precise measurement of the weak mixing angle from the ratio
of neutral current to charged current inclusive cross-sections in
deep-inelastic neutrino-nucleon scattering. The data were gathered at the CCFR
neutrino detector in the Fermilab quadrupole-triplet neutrino beam, with
neutrino energies up to 600 GeV. Using the on-shell definition,
${\rm sin ^2\theta_W} \equiv 1 - \frac{{\rm M_W} ^2}{{\rm M_Z} ^2}$,
we obtain
$
{\rm sin ^2\theta_W} = 0.2218
\pm 0.0025 ({\rm stat.}) \pm 0.0036 ({\rm exp.\: syst.})
\pm 0.0040 ({\rm model})
$.
}

\noindent
PACS numbers:  13.15.Jr, 12.15.Mm, 14.80.Er

\renewcommand\baselinestretch{1.3}

  The standard model (SM) of elementary particle physics describes the
unification of the electromagnetic and weak interactions in terms of a weak
mixing angle, ${\rm sin ^2\theta_W}$.  In the on-shell convention\cite{WMA
definitions}, the mixing angle is defined in terms of the W and Z boson masses:
\begin{equation}
               {\rm sin ^2\theta_W} \equiv
               1 - \frac{{\rm M_W}^2}{{\rm M_Z} ^2}.
\end{equation}
This Letter presents a SM extraction of ${\rm sin ^2\theta_W}$ from the ratio
of neutral current (NC) to charged-current (CC) total cross-sections in
deep-inelastic neutrino-nucleon ($\nu {\rm N}$) scattering,
\begin{equation}
\nu_\mu + {\rm nucleon} \rightarrow \nu_\mu + {\rm hadrons} \;\; ({\rm NC}),
\end{equation}
\begin{equation}
\nu_\mu +{\rm nucleon} \rightarrow \mu^{-} + {\rm hadrons} \; \; ({\rm CC}).
\end{equation}

   The SM predicts that all electroweak processes may be described at lowest
order in perturbation theory by just three independent experimental parameters.
These may be chosen to be the electromagnetic fine structure constant
($\alpha$), the Fermi coupling constant (${\rm G_F}$) and the mass of the Z
boson (${\rm M_Z}$), all of which have been measured to better than 1 part in
$10^4$. Electroweak processes cannot yet be predicted to this level of accuracy
because higher order perturbative corrections for each process bring in
additional dependence on the masses of the undiscovered top quark (${\rm
M_{top}}$) and, to a lesser extent, the Higgs boson (${\rm M_{Higgs}}$). Within
the SM, the experimental determination of ${\rm sin ^2\theta_W}$
from $\nu {\rm N}$ scattering has very little dependence on ${\rm M_{top}}$ or
${\rm M_{Higgs}}$\cite{MT and MH indep}; in contrast, the SM prediction of
${\rm sin ^2\theta_W}$ from $\alpha$, ${\rm G_F}$ and ${\rm M_Z}$ depends
strongly on ${\rm M_{top}}$. Requiring the ${\rm sin ^2\theta_W}$ from
$\nu{\rm N}$ scattering to agree with the prediction using ${\rm M_Z}$ sets
limits on ${\rm M_{top}}$ which are comparable with the best determinations
from Z and W decay experiments at colliders\cite{LEP fit}. From a more general
perspective, the consistency of the ${\rm M_{top}}$ determinations from
different processes constrains possible physics processes beyond the SM.
Neutrino-nucleon scattering is uniquely sensitive to some proposed models with
an extended Higgs sector or with extra Z's\cite{Langacker}. Comparing the SM
prediction for ${\rm M_W}$ from $\nu {\rm N}$ scattering with the direct
measurements at hadron colliders is a further test of the SM which is almost
independent of ${\rm M_{top}}$ and ${\rm M_{Higgs}}$.

   The E770 event sample of 3.1 million raw event triggers was collected in
1987-8 using the quadrupole-triplet neutrino beam-line at Fermilab. The data
sample for the ${\rm sin ^2\theta_W}$ analysis contained 475627 events after
all cuts, with a mean neutrino energy of 161 GeV and a mean 4-momentum transfer
squared, ${\rm Q}^2 = 36\: {\rm GeV}^2 \! /\! c^2$ . This represents
approximately four times the statistics and almost twice the mean energy and
${\rm Q}^2$ of the most precise previous ${\rm sin ^2\theta_W}$ determinations
from $\nu {\rm N}$ scattering\cite{CDHS,CHARM}.

   The CCFR detector\cite{WKS calibration,BJK calibration} consists of a
neutrino target/calorimeter followed by a muon spectrometer. The muon
spectrometer was not used directly in the ${\rm sin ^2\theta_W}$ analysis. The
target comprises 168 iron plates, each 3 m x 3 m x 5.1 cm, interspersed with 84
liquid scintillation counters (every 10 cm of iron) and 42 drift chambers, each
with x and y planes. It is 17.7m long, weighs 695 metric tons and has a mean
density of $4.2\; {\rm g/cm^{3}}$.

   Both CC and NC interactions initiate a cascade of hadrons in the target that
is registered by the drift chambers and scintillation counters. The muon
produced in CC interactions typically penetrates well beyond the end of the
hadron shower, appearing as a track of drift chamber hits with deposits of
characteristic minimum-ionizing energies in the scintillation counters.
We define the event length, ${\rm L}$, to be the number of scintillation
counters spanned by the event, where the longitudinal event vertex is defined
to be the more upstream of the first 2 consecutive counters with more than 4
times the mean energy deposit of
minimum ionizing muons (``mip's'')\cite{WKS calibration},
and the event end is the counter above the next downstream gap of
3 counters with energies below 0.25 mip's. The mean position of the hits in the
drift chambers immediately downstream from the vertex determines the transverse
vertex coordinates. A calorimetric energy, ${\rm E_{cal}}$, is calculated by
summing up energy deposits in the 20 counters immediately downstream from the
vertex. We require the event vertex to be more than 5 counters from the
upstream end of the target and 34 counters from the downstream end and less
than 76.2 cm from the detector center-line. Requiring ${\rm E_{cal}}>
30\;{\rm GeV}$ ensures complete efficiency of the energy deposition trigger.

  The presence of a penetrating muon in CC interactions permits
an approximate partition of CC and NC events by event length:
\begin{equation}
{\rm
{\rm R_{meas}} \; \equiv \;
\frac
{No.\; events\; with\; L\; \leq\; 30\; counters}
{No.\; events\; with\; L\; >\; 30\; counters}
\; \approx\;
\frac
{No.\; NC}
{No.\; CC}
},
\end{equation}
\noindent where ${\rm L} > 30$ counters implies a penetration greater than
about 3.1m of iron. This experimental quantity was translated into a SM value
for ${\rm sin ^2\theta_W}$ using a detailed Monte Carlo-based computer
simulation (MC) of the experiment which modeled the integrated neutrino fluxes,
the relevant physics processes and the response of the CCFR detector. Predicted
values for ${\rm R_{meas}}$ were obtained by generating samples of simulated
events and passing them through the same analysis procedure as the E770 data.
The experimental value of ${\rm sin ^2\theta_W}$ was defined to be the input
value to the MC which returned the same ${\rm R_{meas}}$ as the E770 data. The
relationship between ${\rm R_{meas}}$ and ${\rm sin ^2\theta_W}$ predicted by
the MC is found, in a linear approximation, to be
   $ {\rm sin ^2\theta^{\rm MC}_W} =
                  0.2218 - 1.73\, ( {\rm R^{MC}_{meas}} - 0.4508 )$.
Our experimental determination, ${\rm R_{meas}}$ = 147795/327832 = 0.4508,
corresponds to
\begin{equation}
{\rm sin ^2\theta_W}\; =\; 0.2218\; \pm\; 0.0025 ({\rm stat.})\; \pm\;
0.0036 ({\rm exp.\: syst.})\; \pm\; 0.0040 ({\rm theor.}).
\end{equation}
The experimental and theoretical uncertainties were obtained from the MC by
varying the model parameters within errors, and are itemized in Table 1.

While the analysis presented here is performed completely within the context
of the standard model, our value of $\sin ^2\theta _W$ can be used with the
MC model to calculate a corrected neutral to charged current event ratio
corresponding to the incident $\nu /\overline{\nu }$ flux\cite{flux}, $%
R=(NC^\nu +NC\overline{^\nu})/(CC^\nu +CC^{\overline{\nu}})=
                       0.3117 \pm 0.0014 ({\rm stat.}) \pm
                       0.0018 ({\rm exp.\: syst.}) \pm 0.0014 ({\rm theor.})$.
This value corresponds to a hadron energy cut of 30 GeV on both CC and NC
events. The event ratio is fully corrected for experimental effects such as
acceptance, smearing, and  the $\nu_{\rm e}$ background but no theoretical
corrections are applied other than an isoscaler correction. For the quantity R,
the theoretical uncertainty is almost entirely due to the longitudinal
structure function, R$_{long}$.
In terms of
$R^{\nu (\overline{\nu })} = {\sigma ^{\nu (\overline{\nu })}_{NC} } /
                             {\sigma ^{\nu (\overline{\nu })}_{CC} }$,
\ $R \approx 0.895R^\nu +0.105R\overline{^\nu }$.
The variation of $R$ with $\sin ^2\theta _W$ is very similar to that of
$R_{meas}$ and is given by $ dR/d\sin ^2\theta _W = -0.565$.

   The integrated $\nu_\mu$ and $\overline{\nu}_\mu$ fluxes
for E770 were determined directly from low hadron energy CC
event samples, normalized to the neutrino total
cross sections\cite{PSA flux}.
The final event sample consisted of 86.4\%
$\nu_\mu$, 11.3\% $\overline{\nu}_\mu$ and 2.3\%
$\nu_{\rm e}$ or $\overline{\nu}_{\rm e}$ interactions.
Errors in the $\nu_\mu$ flux tend to cancel in the ratio ${\rm R_{meas}}$,
but the $\nu_{\rm e}$ flux modeling is more critical because
essentially all $\nu_{\rm e}$ events are short enough to appear in the
numerator of ${\rm R_{meas}}$.
The integrated $\nu_{\rm e}$ flux was modeled using a
Monte Carlo simulation of the neutrino beam-line, with the spectra of
secondaries from the proton target parameterized from
experimental production cross-sections\cite{secondary production}.
Approximately 80\% of the $\nu_{\rm e}$'s in the final data sample
were produced from the Ke3 decay mode of charged kaons, whose
modeling is directly related to
the observed $\nu_\mu$ event spectrum. The next largest contribution
to the $\nu_{\rm e}$ flux was from neutral kaon decays ($\sim$16\%),
with smaller contributions from the decays of D mesons, pions,
muons, $\Lambda$'s and $\Sigma^-$'s.

   The modeling of neutrino-induced events in the detector and resolution
smearing effects on the measured ${\rm L}$, ${\rm E_{cal}}$ and vertex
positions were modeled primarily using neutrino and test beam data events%
\cite{WKS calibration,BJK calibration,punchthru}.  Systematic uncertainties
associated with the hadronic energy measurement ${\rm E_{cal}}$ include
possible small NC/CC shower differences, uncertainties in the muon energy
deposit within the hadron shower, and uncertainties in the resolution
function, e/$\pi $ response, and absolute energy scales obtained from
hadron/electron test beam measurements\cite{WKS calibration,BJK calibration}.
The length uncertainties include those associated with the shower length
parameterizations from test beam measurements\cite{punchthru}, the
longitudinal vertex determination, which has been checked against the vertex
of dimuon events, counter inefficiencies and noise, and effects from
dimuon production. Small differences in the transverse vertex for NC and CC
events due to the muon drift chamber hits were determined from the
analysis of CC events with these hits removed.

   The NC and CC differential cross-sections were modeled using a
QCD-enhanced quark-parton description of the nucleon.
The quark distributions were obtained by using a modified
Buras-Gaemers parameterization\cite{Buras-Gaemers} of
CC nucleon structure
functions measured in the same CCFR experiment\cite{PZQ SF}.
The strange quark component, parameterized by the momentum fraction relative to
the non-strange sea, $\kappa = 2s / (\overline{u} +\overline{d})$,
was determined from an analysis
of CCFR dimuon events\cite{SAR} which arise from the
muonic decays of charm quarks produced in CC scattering
off down and strange quarks and anti-quarks.
The threshold suppression for this process, due to the
mass of the charm quark, was modeled using a leading order
slow-rescaling formalism, with a fitted effective
charm quark mass of
${\rm M_c}\, =\, 1.31\, \pm\, 0.24\: {\rm GeV\! /\! c}^2$ \cite{SAR}.
The level of the charm sea was assumed to be 10\% of the strange sea,
consistent with a wrong-sign muon analysis from a previous CCFR neutrino
experiment using the same detector\cite{wrong-sign}. Our parameterization of
the R$_{long}$ is based on QCD predictions and data from
charged lepton scattering experiments\cite{Whitlow}. A correction for the
difference between u and d valence quark distributions in nucleons, obtained
from muon scattering data\cite{NMC}, was applied to account for the 5.67\%
excess of neutrons over protons in the target. Radiative corrections to the
scattering cross-sections were applied using computer code supplied by
Bardin\cite{Bardin},
assuming values of ${\rm M_{top}}\, =\, 150\: {\rm GeV\! /\! c}^2$
and ${\rm M_{Higgs}}\, =\, 100$ ${\rm GeV\! /\! c}^2$.

   Figure 1 shows the length distribution of the E770 final data sample and a
MC simulated event sample. Events reaching the muon spectrometer, comprising
79\% of the CC interactions, have been left out for clarity but are included in
the normalization of the MC event sample to the data. The remaining CC events
have a muon which either has a low energy and ranges out in the neutrino target
or has a large opening angle with respect to the incident neutrino and exits
through the side of the target. The production energy and angular distributions
of these muons are very well constrained by the CCFR structure function
measurements, and their propagation through the target has been precisely
parameterized using large samples of muons from test beam and neutrino data.
The events with length less than or equal to the 30 counter partition of
equation 4 are predominantly true $\nu_\mu$ (or $\overline{\nu}_\mu$)
NC events, with 22.9\% and 7.3\% backgrounds from short $\nu_\mu$ and
$\overline{\nu}_\mu$ CC events and $\nu_{\rm e}$ events, respectively. Since
the NC and $\nu_{\rm e}$ event lengths fall well short of the 30 counter
partition, the sensitivity to the modeling of the hadron shower length is
minimal.  The good agreement between data and MC for event lengths greater
than 25 counters reinforces confidence in the estimate of the CC component of
the event sample in the `NC' length region of 30 counters or less.
In addition, the data and
MC distributions in ${\rm E_{cal}}$ and vertex radial position also agree well.

    The most precise previous determinations of
${\rm sin ^2\theta_W}$ in $\nu {\rm N}$ are from
the CDHS\cite{CDHS} and CHARM\cite{CHARM} collaborations.
After adjusting to our theoretical assumptions for
the charm quark mass,
${\rm M_c} = 1.31 \pm 0.24$ ${\rm GeV\! /\! c}^2$ \cite{SAR},
and top quark mass,
${\rm M_{top}}\, =\, 150\: {\rm GeV\! /\! c}^2$,
these experiments yield measurements of
${\rm sin ^2\theta_W} = 0.2225 \pm 0.0066$ (CDHS)
and
${\rm sin ^2\theta_W} = 0.2319 \pm 0.0065$ (CHARM),
in agreement with our result,
${\rm sin ^2\theta_W} = 0.2218 \pm 0.0059$.

    Combining our value of ${\rm sin ^2\theta_W}$ with the
precise measurement of the Z boson mass,
${\rm M_Z} = 91.187 \pm 0.007\: {\rm GeV\! /\! c}^2 $ \cite{Z mass},
gives ${\rm M_W} = 80.44 \pm 0.31$ ${\rm GeV\! /\! c}^2$,
and corresponds in the SM to
   ${\rm M_{top}}   = 190 ^{+39+12}_{-48-14}\; {\rm GeV\! /\!c^2}$
\cite{Langacker comm.},
where the central value and first set of uncertainties assume
${\rm M_{Higgs}} = 300\; {\rm GeV\! /\!c^2}$ and the second set of
uncertainties come from varying ${\rm M_{Higgs}}$
between 60 and 1000 ${\rm GeV\! /\!c^2}$.  Our results are consistent
with the values,
${\rm M_W} = 80.25 \pm 0.10$ ${\rm GeV\! /\! c}^2$
and
${\rm M_{top}}   = 166 ^{+17+19}_{-19-22}\; {\rm GeV\! /\!c^2}$,
from a SM fit to a large number of experimental results
from Z decays at the LEP electron collider\cite{LEP fit},
and our ${\rm M_W}$ value also agrees with
the world average measured value from hadron
colliders, ${\rm M_W} = 80.14 \pm 0.27$ ${\rm GeV\! /\! c}^2$
\cite{MW}.

   In summary, our value is the most precise determination
of the on-shell ${\rm sin ^2\theta_W}$ in a single experiment and is
consistent with previous determinations in
$\nu {\rm N}$ scattering and other processes.

   We thank the management and staff of Fermilab, and acknowledge
the help of many individuals at our home institutions. This
research was supported by the National Science Foundation
and the Department of Energy. SRM acknowledges the support
of the Alfred P. Sloan Foundation.

\vfill\eject

\parskip 0.01in

\bibliographystyle{unsrt}

\begin{thebibliography}{99}
\bibitem{flux}  Approximate parameterizations of the neutrino energy spectra
of the muon neutrino and antineutrino events as a function of neutrino
energy $E$ (in GeV) are, with arbitrary overall normalization,  $N_\nu
(E)$ = 1.000${\rm e}^{-\frac{(E-84)^2}{2383}}$ + 0.057${\rm e}^{-\frac{%
(E-160)^2}{5046}}$ + 0.154${\rm e}^{-\frac{(E-266)^2}{1472}} + $0.139${\rm
e}^{-
\frac{(E-232)^2}{29246}}$ and $N_{\bar \nu }(E)$ = 0.145${\rm e}^{-\frac{%
(E-50)^2}{246}}$ + 0.376${\rm e}^{-\frac{(E-86)^2}{1749}}$ + 0.028${\rm e}^{-
\frac{(E-263)^2}{2287}}$ + 0.087${\rm e}^{-\frac{(E-49)^2}{36532}}.$
\end{thebibliography}

\vfill\eject
\clearpage
\newpage

\vskip .2in

{\small

 \begin{table}

{ {\sc TABLE 1}. Uncertainties in the measurement of ${\rm sin
^2\theta_W}$.}

\vspace{0.3 cm}

\begin{tabular}{|r|c|}
\hline
 { data statistics } & {  0.0024} \\
 { Monte Carlo statistics } & { 0.0006} \\ \hline
 { TOTAL STATISTICS \hfill } & { 0.0025} \\ \hline\hline
  { muon neutrino flux} & { 0.0005} \\
 ($\nu _e\: \pm$ 4.2\% )\ \  { electron neutrino flux } & { 0.0023 } \\
  { transverse vertex} & { 0.0009} \\
  ($\pm 25\%$) { cosmic ray subtraction} & { 0.0003}\\
  { Energy Measurement  \hfill } &  \\
  { NC/CC shower difference} & { 0.0007} \\
  { muon energy loss in shower region} & { 0.0005} \\
  ($\pm 10\%$) { hadron energy resolution} & { 0.0005} \\
  ($\pm 1\%$) { absolute energy scale} & { 0.0018} \\
  { Event Length     \hfill } &  \\
  { hadron shower length} & { 0.0010} \\
  ($\pm 2.5$ cm) { vertex determination} & { 0.0010} \\
  { counter efficiency and noise} & { 0.0009} \\
  { dimuon production} & { 0.0003} \\
  \hline
  { TOTAL EXP. SYST. \hfill } & { 0.0036 } \\ \hline\hline
  (${\rm M_c}\, =\, 1.31\,\pm\, 0.24\;{\rm GeV\! /\! c}^2$)
                                    \ \ { charm prod.  } &
                                                       { 0.0030}  \\
 ( R$ _{long} \mp $ 15\% )\ \ { long. SF } & { 0.0019} \\
  (${\rm C/S}\, =\, 0.10 \pm 0.15$)\ \ { charm sea } & { 0.0015} \\
  { rad. corrections  } & { 0.0007}  \\
   { higher twist } & { 0.0005} \\
   { non-isoscalar target} & { 0.0004 } \\
 ($\kappa = 0.37 \mp 0.05$)\ \ { strange sea } & { 0.0003} \\
   { structure functions } & { 0.0003} \\ \hline
  { TOTAL PHYSICS MODEL \hfill } & { 0.0040 } \\  \hline
\end{tabular}
 \end{table}
}

\vfill\eject
\clearpage
\newpage

\centerline{ Figure Captions}

\vskip .3in

\noindent
{ {\sc FIGURE 1}. Data and Monte Carlo (MC) event length distributions.
The data are represented by dots and the MC prediction
by the solid line. Also shown are the MC contributions from
NC $\nu_\mu$ events (``NC"), CC $\nu_\mu$ events (``CC") and
combined NC and CC interactions from
$\nu_{\rm e}$ or $\overline{\nu}_{\rm e}$ (``$\nu_{\rm e}$").
The inset shows the data, total MC and the NC contribution to the
MC for the region L$\geq$25 counters.
}

\vskip .1in

\vfill\eject
\clearpage
\newpage

\end{document}